\documentclass{article}

\usepackage{graphicx}

\title{New model for the neutrino mass matrix}

\author{L.\ Lavoura \\
\small Universidade T\'ecnica de Lisboa \\
\small Centro de F\'\i sica das Interac\c c\~oes Fundamentais \\
\small Instituto Superior T\'ecnico, 1049-001 Lisboa, Portugal}

\date{31 May 2000}

\begin{document}
\maketitle

\begin{abstract}
I suggest a model based on a softly broken symmetry $L_e - L_\mu - L_\tau$
and on Babu's mechanism for two-loops radiative generation
of the neutrino masses.
The model predicts that one of the physical neutrinos ($\nu_3$) is massless
and that its component along the $\nu_e$ direction ($U_{e3}$) is zero.
Moreover,
if the soft-breaking term is assumed to be very small,
then the vacuum oscillations of $\nu_e$ have almost maximal amplitude
and solve the solar-neutrino problem.
New scalars are predicted in the $10\, {\rm TeV}$ energy range,
and a breakdown of $e$--$\mu$--$\tau$ universality
should not be far from existing experimental bounds.
\end{abstract}

\vspace{6mm}

In a model without right-handed (singlet) neutrinos,
the three weak-interaction-eigenstate neutrinos $\nu_e$,
$\nu_\mu$,
and $\nu_\tau$ may acquire $\left| \Delta I \right| = 1$ Majorana masses
given by the following term in the Lagrangian:
\begin{equation}
{\cal L}_{\rm mass}^{(\nu)} =
{\textstyle \frac{1}{2}}
\left( \begin{array}{ccc}
\nu_e^T & \nu_\mu^T & \nu_\tau^T
\end{array} \right)
C^{-1}
{\cal M}
\left( \begin{array}{c}
\nu_e \\ \nu_\mu \\ \nu_\tau
\end{array} \right)
- 
{\textstyle \frac{1}{2}}
\left( \begin{array}{ccc}
\overline{\nu_e} & \overline{\nu_\mu} & \overline{\nu_\tau}
\end{array} \right)
C
{\cal M}^\ast
\left( \begin{array}{c}
\overline{\nu_e}^T \\ \overline{\nu_\mu}^T \\ \overline{\nu_\tau}^T
\end{array} \right).
\label{mass term}
\end{equation}
Here,
$C$ is the Dirac--Pauli charge-conjugation matrix
and $\cal M$ is a $ 3 \times 3$ symmetric mass matrix.
One may diagonalize $\cal M$ with help of a unitary matrix $U$
in the following way:
\begin{equation}
U^T {\cal M} U = {\rm diag} \left( m_1, m_2, m_3 \right),
\label{diagonalization}
\end{equation}
where $m_1$,
$m_2$,
and $m_3$ are real and non-negative.
The physical neutrinos $\nu_1$,
$\nu_2$,
and $\nu_3$ are given by
\begin{equation}
\left( \begin{array}{c} \nu_e \\ \nu_\mu \\ \nu_\tau \end{array} \right)
= U \left( \begin{array}{c} \nu_1 \\ \nu_2 \\ \nu_3 \end{array} \right).
\label{physical neutrinos}
\end{equation}
Then,
\begin{equation}
{\cal L}_{\rm mass}^{(\nu)} =
{\textstyle \frac{1}{2}} \sum_{i=1}^3 m_i \left(
\nu_i^T C^{-1} \nu_i - \overline{\nu_i} C \overline{\nu_i}^T \right).
\label{mass term diagonalized}
\end{equation}

Experiment indicates that two linearly independent
squared-mass differences among the three physical neutrinos
differ by a few orders of magnitude.
Indeed,
$\Delta m^2_{\rm atm}$ is of order $10^{-3}\, {\rm eV}^2$,
while $\Delta m^2_\odot$ may be either of order $10^{-5}\, {\rm eV}^2$,
in the case of the MSW solution for the solar-neutrino puzzle,
or of order $10^{-10}\, {\rm eV}^2$,
in the case of the vacuum-oscillations (``just so'') solution.
It is customary to identify $\nu_3$ as the neutrino
which has a mass much different from the masses of the other two,
{\it viz.},
\begin{equation}
\left| m_2^2 - m_1^2 \right| = \Delta m^2_\odot \ll
\left| m_3^2 - m_1^2 \right| \approx \left| m_3^2 - m_2^2 \right|
\approx \Delta m^2_{\rm atm}.
\label{mass hierarchy}
\end{equation}
Then,
the negative result of CHOOZ's search for $\nu_e$ oscillations \cite{CHOOZ}
is interpreted as $\left| U_{e3} \right| \le 0.217$,
which is valid for $\Delta m^2_{\rm atm} \ge 2 \times 10^{-3}\, {\rm eV}^2$.

It has been pointed out \cite{barbieri} that
the assumption of an approximate lepton-number symmetry
$\bar L \equiv L_e - L_\mu - L_\tau$
(where $L_e$ is the electron number,
$L_\mu$ is the muon number,
and $L_\tau$ is the tau number)
may constitute a good starting point for a model of the neutrino mass matrix.
Indeed,
if there are no $\left| \Delta \bar L \right| = 2$ mass terms then
\begin{equation}
{\cal M} = \left( \begin{array}{ccc}
0 & r b & b \\ r b & 0 & 0 \\ b & 0 & 0
\end{array} \right),
\label{M with Lbar}
\end{equation}
where $b$ and $r$ may,
without loss of generality,
be taken to be real and positive.
The mass matrix in Eq.~(\ref{M with Lbar}) yields $m_3 = 0$,
$m_1 = m_2 = b \sqrt{1 + r^2}$,
and
\begin{equation}
U = \left( \begin{array}{ccc}
{\displaystyle \frac{1}{\sqrt{2}}} &
- {\displaystyle \frac{i}{\sqrt{2}}} &
0 \\*[3mm]
{\displaystyle \frac{r}{\sqrt{2 \left( 1 + r^2 \right)}}} &
{\displaystyle \frac{i r}{\sqrt{2 \left( 1 + r^2 \right)}}} &
{\displaystyle \frac{1}{\sqrt{1 + r^2}}} \\*[3mm]
{\displaystyle \frac{1}{\sqrt{2 \left( 1 + r^2 \right)}}} &
{\displaystyle \frac{i}{\sqrt{2 \left( 1 + r^2 \right)}}} &
- {\displaystyle \frac{r}{\sqrt{1 + r^2}}}
\end{array} \right).
\label{U with Lbar}
\end{equation}
This is good for the following reasons:
\begin{enumerate}
\item The negative result of CHOOZ's search for $\nu_e$ oscillations
gets explained through $U_{e3} = 0$.
\item Since $4 \left| U_{e1} U_{e2} \right|^2 = 1$,
vacuum oscillations of $\nu_e$ with maximal amplitude
would occur were $m_1 \neq m_2$,
opening way for the ``just so'' solution of the solar-neutrino problem
to apply.
\item It is intuitive to expect $r$ to be close to $1$.
Now,
if $r = 1$ then $\nu_\mu$--$\nu_\tau$ mixing is maximal,
and this explains the atmospheric-neutrino anomaly.
\end{enumerate}

On the other hand,
$\bar L$ must be broken,
because $m_1 = m_2$ does not allow
for oscillations between $\nu_1$ and $\nu_2$
and a solution of the solar-neutrino puzzle.
A good choice,
in order to avoid unpleasant majorons,
would be to have $\bar L$ to be softly broken;
this would moreover permit a natural explanation for
$\Delta m^2_\odot \ll \Delta m^2_{\rm atm}$.
This option has been suggested by Joshipura and Rindani \cite{joshipura};
however,
in those authors' models
there is no predictive power for the form of the mixing matrix $U$,
a fact which impairs the immediate interest
and experimental testability of those models.

In this paper I put forward a simple model
with softly broken $\bar L$ which maintains some predictive power.
The model is based on Babu's mechanism
for two-loops radiative generation of the neutrino masses \cite{babu}.
I remind that,
in general,
Babu's mechanism leads to one neutrino remaining massless;
however,
whereas that general mechanism cannot predict the $\nu_e$,
$\nu_\mu$,
and $\nu_\tau$ components of the massless neutrino,
the specific model that I shall put forward
retains the exact-$\bar L$ prediction $U_{e3} = 0$.
Moreover,
in my model there is a rationale
for the $\nu_e$ oscillations of maximal amplitude,
and for the tiny mass difference $\Delta m^2_\odot$,
which allow a ``just so'' explanation of the solar-neutrino deficit;
that rationale is provided by the naturalness of the assumption
that the term which breaks $\bar L$ softly is very small.

In my model I just introduce in the scalar sector,
above and beyond the usual standard-model doublet
$\phi = \left( \begin{array}{cc} \varphi^+ & \varphi^0 \end{array} \right)^T$,
one singly-charged singlet $f^+$ with $\bar L = 0$,
together with two doubly-charged singlets $g^{2+}$ and $h^{2+}$,
and their Hermitian conjugates.
The difference between $g^{2+}$ and $h^{2+}$ lies in that
the former field has $\bar L = 0$
whereas $h^{2+}$ has $\bar L = -2$.
The Yukawa couplings of the leptons are $\bar L$-invariant
and are given by
\begin{eqnarray}
{\cal L}_{\rm Y}^{\rm (l)} &=&
- \frac{m_e}{v}
\left( \begin{array}{cc}
\overline{\nu_{eL}} & \overline{e_L} \end{array} \right)
\left( \begin{array}{c}
\varphi^+ \\ \varphi^0 \end{array} \right)
e_R
- \frac{m_\mu}{v}
\left( \begin{array}{cc}
\overline{\nu_{\mu L}} & \overline{\mu_L} \end{array} \right)
\left( \begin{array}{c}
\varphi^+ \\ \varphi^0 \end{array} \right)
\mu_R
- \frac{m_\tau}{v}
\left( \begin{array}{cc}
\overline{\nu_{\tau L}} & \overline{\tau_L} \end{array} \right)
\left( \begin{array}{c}
\varphi^+ \\ \varphi^0 \end{array} \right)
\tau_R
\nonumber\\*[1mm]
 & &
+ f^+ \left[
f_\mu \left(
\nu_{eL}^T C^{-1} \mu_L - e_L^T C^{-1} \nu_{\mu L}
\right)
+
f_\tau \left(
\nu_{eL}^T C^{-1} \tau_L - e_L^T C^{-1} \nu_{\tau L}
\right)
\right]
\nonumber\\*[2mm]
 & &
+ e_R^T C^{-1} \left[ g^{2+} \left( g_\mu \mu_R + g_\tau \tau_R \right)
+ h^{2+} h_e e_R \right] + {\rm H.c.},
\label{Yukawa couplings}
\end{eqnarray}
where $f_\mu$,
$f_\tau$,
$g_\mu$,
$g_\tau$,
and $h_e$ are complex coupling constants.
Notice that,
in the first line of Eq.~(\ref{Yukawa couplings}),
I have already taken,
without loss of generality,
the Yukawa couplings of $\phi$ to be flavor-diagonal;
$v$ denotes the vacuum expectation value of $\varphi^0$.

The scalar potential $V$ has a trivial part,
$V_{\rm trivial}$,
which is a quadratic polynomial in $\phi^\dagger \phi$,
$f^- f^+$,
$g^{2-} g^{2+}$,
and $h^{2-} h^{2+}$.
Besides,
$V$ includes two other terms,
with complex coefficients $\lambda$ and $\epsilon$:
\begin{equation}
V = V_{\rm trivial} + \left( \lambda f^- f^- g^{2+} + \epsilon g^{2-} h^{2+}
+ {\rm H.c.} \right). 
\label{potential}
\end{equation}
The term with coefficient $\epsilon$ breaks $\bar L$ softly.
I make the following assumptions:
this is the only $\bar L$-breaking term in the theory,
and $\epsilon$ is small.
These assumptions are technically natural
in the sense of 't Hooft \cite{hooft}.\footnote{Notice that
the possible $\bar L$-breaking term $f^- f^- h^{2+}$ has dimension
higher than the one of $g^{2-} h^{2+}$,
and therefore the assumption of its absence is natural.}

From now on I shall assume,
without loss of generality,
$f_\mu$,
$f_\tau$,
$g_\tau$,
$h_e$,
$\lambda$,
and $\epsilon$ to be real and positive.
Only $g_\mu$ remains,
in general,
complex.

The neutrino mass term ${\cal M}_{e \mu}$ does not break $\bar L$
and is generated at two-loops level by the Feynman diagram in Figure 1.
A similar diagram generates ${\cal M}_{e \tau}$.
In both cases,
there is in the diagram
an inner charged lepton which may be either $\mu$ or $\tau$.
It is clear that the mass terms thus generated obey the relation
\begin{equation}
r \equiv \frac{{\cal M}_{e \mu}}{{\cal M}_{e \tau}}
= \frac{f_\mu}{f_\tau}.
\label{mass relation}
\end{equation}
Contrary to what happens in Zee's model \cite{zee},
this ratio of mass terms is not proportional
to a ratio of squared charged-lepton masses \cite{jarlskog}.
As seen before,
in order to obtain maximal $\nu_\mu$--$\nu_\tau$ mixing
one would like to have $r \approx 1$.
In the present model,
this means that the coupling constants $f_\mu$ and $f_\tau$
should be approximately equal.
In Zee's model,
on the other hand,
one winds up with the rather unrealistic constraint
$f_\mu / f_\tau \approx \left( m_\tau / m_\mu \right)^2$.

Let us check whether the diagram in Figure 1 is able to yield
neutrino masses of the right order of magnitude.
As we shall see later,
we would like to obtain $\left| {\cal M}_{e \mu} \right| \approx
\left| {\cal M}_{e \tau} \right| \approx \sqrt{\Delta m^2_{\rm atm}}
\sim 10^{-2}$--$10^{-1}\, \rm{eV}$.
Now,
from the diagram in Figure 1 with an inner $\tau$ one obtains
\begin{equation}
{\cal M}_{e \mu} = - 2 \lambda f_\mu f_\tau g_\tau m_e m_\tau
\frac{I}{\left( 16 \pi^2 \right)^2},
\label{matrix element}
\end{equation}
where
\begin{eqnarray}
I &=& \frac{1}{\pi^4}
\int \! d^4 k\, \frac{1}{k^2 - m_f^2}\, \frac{1}{k^2 - m_e^2}\,
\int \! d^4 q\,
\frac{1}{q^2 - m_f^2}\,
\frac{1}{q^2 - m_\tau^2}\,
\frac{1}{\left( k-q \right)^2 - m_g^2}
\label{integral1} \\
  &=& \frac{1}{2 \left( m_f^2 - m_\tau^2 \right)} \int_0^\infty
\frac{dy}{\left( y + 1 \right) \left( y + x_e \right)}
\left[ p \ln \frac{y + x_g + 1 + p}{y + x_g + 1 - p}
- p^\prime \ln
\frac{y + x_g + x_\tau + p^\prime}{y + x_g + x_\tau - p^\prime} \right.
\nonumber \\
  & &
\left.
+ \left( 1 - x_\tau \right) \ln x_g
+ \left( x_\tau - x_g - y \right) \ln x_\tau
\right].
\label{integral2}
\end{eqnarray}
Here,
$x_e = m_e^2 / m_f^2$,
$x_\tau = m_\tau^2 / m_f^2$,
$x_g = m_g^2 / m_f^2$,
and
\begin{eqnarray}
p &=& \sqrt{\left( y + x_g - 1 \right)^2 + 4 y},
\label{k}
\\
p^\prime &=& \sqrt{\left( y + x_g - x_\tau \right)^2 + 4 y x_\tau}.
\label{kprime}
\end{eqnarray}
The integral in Eq.~(\ref{integral2}) is convergent
and may be computed numerically.\footnote{It is not possible
to use the approximations $m_e = m_\tau = 0$
because they lead to infrared divergences.
This is not a problem,
since those divergences are logarithmic
and ${\cal M}_{e \mu}$ in Eq.~(\ref{matrix element})
also includes a factor $m_e m_\tau$.}
For $m_e$,
$m_\tau \ll m_f$ and $m_g \approx m_f$,
one finds $I$ to be of order $m_f^{-2}$.

In my estimate of ${\cal M}_{e \mu}$
I shall therefore set $I \approx m_f^{-2}$.
The bounds from $e$--$\mu$--$\tau$ universality
in $\mu$ decay and in $\tau$ decay
are $f_\mu / m_f\,
\raise.3ex\hbox{$<$\kern-.75em\lower1ex\hbox{$\sim$}}\,
10^{-4}\, {\rm GeV}^{-1}$
and
$f_\tau / m_f\,
\raise.3ex\hbox{$<$\kern-.75em\lower1ex\hbox{$\sim$}}\,
10^{-4}\, {\rm GeV}^{-1}$ \cite{babu};
if one allows $f_\mu f_\tau / m_f^2$
to be as high as $10^{-8}\, {\rm GeV}^{-2}$,
then one obtains
\begin{equation}
\left| {\cal M}_{e \mu} \right| \approx 10^{-15} \lambda g_\tau.
\end{equation}
It is reasonable to assume that the Yukawa coupling $g_\tau$
is of the same order of magnitude as the Yukawa couplings
$f_\mu$ and $f_\tau$,
and that the dimensionful scalar-potential coupling constant $\lambda$
is of the same order of magnitude as both $m_f$ and $m_g$.
This leads to $g_\tau / \lambda \sim f_\mu / m_f
\sim 10^{-4}\, {\rm GeV}^{-1}$.
Fortunately the product $\lambda g_\tau$ stays free.
In order to obtain $\left| {\cal M}_{e \mu} \right|
\sim 10^{-2}\, {\rm eV}$ it is then sufficient to assume
\begin{eqnarray}
\lambda \approx m_g \approx m_f &\sim& 10^4\, {\rm GeV},
\label{masses} \\
f_\mu \approx f_\tau \approx g_\tau &\sim& 1.
\label{Yukawas}
\end{eqnarray}
Extra factors of order 1 may easily enhance $\left| {\cal M}_{e \mu} \right|$
and bring it up to the desired value $0.06\, {\rm eV}$.

The assumption,
made in Eq.~(\ref{Yukawas}),
that the Yukawa couplings are of order 1,
may seem unrealistic.\footnote{Notice however that,
in the standard model,
the top-quark Yukawa coupling is also very close to 1.}
However,
there are no experimental indications against this possibility
when the masses of $f^+$ and of $g^{2+}$
are assumed to be as high as $10\, {\rm TeV}$.\footnote{Concerns
about the breakdown of perturbativity
are only justified for Yukawa couplings
$\raise.4ex\hbox{$>$\kern-.75em\lower1.2ex\hbox{$\sim$}} 4 \pi$,
{\it i.e.},
of order 10 or more.}
For instance,
$g^{2+}$ mediates the unobserved decay $\tau^- \to \mu^- e^+ e^-$;
however,
by comparing that decay with the standard
$\tau^- \to \mu^- \bar \nu_\mu \nu_\tau$,
one easily reaches the conclusion that
${\rm BR} \left( \tau^- \to \mu^- e^+ e^- \right)$ should be
at least one order of magnitude below the present experimental bound,
when $m_g \approx 10\, {\rm TeV}$ and $\left| g_\mu g_\tau \right| \approx 1$.
A more complicated process is $e^+ e^- \to \tau^+ \tau^-$,
which is mediated by $g^{2+}$ exchange in the $t$ channel.
The amplitude $A$ for this process is
\begin{eqnarray}
A &=&
\frac{i e^2}{s}
\left[ \bar v(e) \gamma^\mu u(e) \right]
\left[ \bar u(\tau) \gamma_\mu v(\tau) \right]
+ \frac{i e^2}{3 \left( s - m_z^2 \right)}
\left[ \bar v (e) \gamma^\mu \gamma_5 u(e) \right]
\left[ \bar u (\tau) \gamma_\mu \gamma_5 v(\tau) \right]
\nonumber \\
  & &
- \frac{i g_\tau^2}{8 \left( t - m_g^2 \right)}
\left[ \bar v (e) \gamma^\mu \left( 1 + \gamma_5 \right) u(e) \right]
\left[ \bar u (\tau) \gamma_\mu \left( 1 + \gamma_5 \right) v(\tau) \right].
\label{amplitude}
\end{eqnarray}
I have used the convenient approximations $m_e = m_\tau = 0$
and $\sin^2 \theta_w = 1/4$ in writing down the standard-model amplitude,
and a Fierz transformation in the non-standard contribution.
If one defines $j = 2 m_g^2 / s$,
$z = g_\tau^2 / \left( 2 e^2 \right)$,
and $l = 3 \left( s - m_z^2 \right) / s$, 
then one finds
\begin{equation}
\frac{d \sigma}{d \cos \theta} \propto
\frac{l^2 + 1}{l^2} \left( 1 + \cos^2 \theta \right)
+ \frac{4}{l}\, \cos \theta
+ z \frac{l + 1}{l}
\frac{\left( 1 + \cos \theta \right)^2}{1 + j + \cos \theta}
+ z^2 \frac{\left( 1 + \cos \theta \right)^2}
{\left( 1 + j + \cos \theta \right)^2},
\label{cross section}
\end{equation}
where $\theta$ is the angle between the momenta of $e^-$ and of $\tau^-$
in the center-of-momentum frame.
From the differential cross section in Eq.~(\ref{cross section})
one easily checks that the deviations of both the total cross section
and the forward--backward asymmetry from their standard-model predictions
are completely negligible when $m_g \sim 10\, {\rm TeV}$,
even if $g_\tau$ is as large as 1.

Except for ${\cal M}_{e \mu}$ and ${\cal M}_{e \tau}$,
all other matrix elements of ${\cal M}$ break $\bar L$ and,
therefore,
they will all be proportional to the $\bar L$-breaking parameter $\epsilon$,
which is assumed to be small.
The matrix elements ${\cal M}_{\mu \mu}$,
${\cal M}_{\mu \tau}$,
and ${\cal M}_{\tau \tau}$ arise at two loops from the diagram in Figure 2.
In order to obtain a non-zero ${\cal M}_{ee}$ one must go to three loops
and use for instance the diagram in Figure 3.
In that diagram there are two inner charged leptons
which may be either $\mu$ or $\tau$;
therefore,
there is a contribution to  ${\cal M}_{ee}$
proportional to $m_\tau^2$,
and that matrix element should not be neglected
in spite of it only arising at three-loops level.

The diagram in Figure 2
clearly leeds to the following relation:
\begin{equation}
{\cal M}_{\mu \mu} : {\cal M}_{\mu \tau} : {\cal M}_{\tau \tau}
= f_\mu^2 : \left( f_\mu f_\tau \right) : f_\tau^2
= r^2 : r : 1.
\label{second mass relation}
\end{equation}
One thus obtains that in the present model
\begin{equation}
{\cal M} = \left( \begin{array}{ccc}
a & r b & b \\ r b & r^2 c & r c \\ b & r c & c
\end{array} \right),
\label{form of M}
\end{equation}
where $a$,
$b$,
and $c$ are complex numbers with mass dimension,
while $r = f_\mu / f_\tau$ is a real dimensionsless number
which should in principle be of order 1.
The masses $a$ and $c$ are suppressed relative to $b$
by the soft-breaking parameter $\epsilon$.

The mass matrix in Eq.~(\ref{form of M}) immediately leads
to two predictions of this model:
there is one massless neutrino ($\nu_3$)
and its component along the $\nu_e$ direction,
{\it i.e.},
$U_{e3}$,
vanishes.
Indeed,
the diagonalizing matrix $U$ reads
\begin{equation}
U = \left( \begin{array}{ccc}
\cos \psi & - i \sin \psi & 0 \\*[2mm]
e^{i \alpha} {\displaystyle \frac{r \sin \psi}{\sqrt{1 + r^2}}} &
e^{i \alpha} {\displaystyle \frac{i r \cos \psi}{\sqrt{1 + r^2}}} &
{\displaystyle \frac{1}{\sqrt{1 + r^2}}} \\*[3mm]
e^{i \alpha} {\displaystyle \frac{\sin \psi}{\sqrt{1 + r^2}}} &
e^{i \alpha} {\displaystyle \frac{i \cos \psi}{\sqrt{1 + r^2}}} &
- {\displaystyle \frac{r}{\sqrt{1 + r^2}}}
\end{array} \right) .\
{\rm diag} \left( e^{i \theta_1}, e^{i \theta_2}, 1 \right),
\label{form of U}
\end{equation}
{\it cf.} Eq.~(\ref{U with Lbar}).
In the matrix of Eq.~(\ref{form of U})
$\alpha \equiv
\arg \left[ a b^\ast + b c^\ast \left( 1 + r^2 \right ) \right]$
is a physically meaningless phase.
The Majorana phases $\theta_1$ and $\theta_2$
are necessary in order to obtain real and positive $m_1$ and $m_2$.
The sole physically observable phase
is $2 \left( \theta_1 - \theta_2 \right)$ \cite{livro}.
The mixing angle $\psi$ is given by
\begin{equation}
\tan \psi = \sqrt{1 + \varepsilon^2} + \varepsilon,
\label{psi}
\end{equation}
where
\begin{equation}
\varepsilon = \frac
{\left| c \right|^2 \left( 1 + r^2 \right) - \left| a \right|^2}
{2 \sqrt{1 + r^2} \left| a b^\ast + b c^\ast \left( 1 + r^2 \right) \right|}
\label{varepsilon}
\end{equation}
is a parameter of order $\epsilon$,
just as $a/b$ and $c/b$,
and may therefore be assumed to be very small.
Thus,
$\psi$ is close to $45^\circ$.
The amplitude of the vacuum oscillations of $\nu_e$
relevant for the solution of the solar-neutrino problem
is $4 \left| U_{e1} U_{e2} \right|^2 = \left( 1 + \varepsilon^2 \right)^{-1}$,
{\it i.e.},
almost maximal.
Thus,
the present model favors a ``just so'' solution of the solar-neutrino puzzle.

The soft-breaking parameter $\epsilon$ should be tiny.
Indeed,
one finds
\begin{equation}
\frac{\Delta m^2_\odot}{\Delta m^2_{\rm atm}}
\approx 2 \frac{\left| a b^\ast + b c^\ast \left( 1 + r^2 \right) \right|}
{\left| b \right|^2 \sqrt{1 + r^2}} \sim \epsilon;
\end{equation}
as we want the ``just so'' solution for the solar-neutrino puzzle to apply,
we must accept $\epsilon$ to be of order $10^{-7}$.
Such a tiny soft breaking of $\bar L$ may eventually be explained
by some new physics at a very high energy scale.

From the non-observation of neutrinoless double beta decay
one derives the bound
$\left| {\cal M}_{ee} \right| \leq 0.2\, {\rm eV}$ \cite{double beta}.
This is not a problem to the present model.
Indeed,
as $m_3$ is predicted to vanish,
$m_1$ and $m_2$ should both be very close to
$\sqrt{\Delta m^2_{\rm atm}} \approx 0.06\, {\rm eV}$.
Thus,
in the approximation $\cos^2 \psi = \sin^2 \psi = 1/2$,
one has
\begin{equation}
\left| {\cal M}_{ee} \right| \approx \left( 0.03\, {\rm eV} \right)
\left| e^{2 i \left( \theta_1 - \theta_2 \right)} - 1 \right|
< 0.2\, {\rm eV}.
\label{Mee}
\end{equation}
Moreover,
the phase $2 \left( \theta_1 - \theta_2 \right)$
is very close to zero---indeed,
it vanishes in the limit of $\bar L$ conservation.

In conclusion,
the model that I have presented in this paper
makes the exact predictions $m_3 = 0$ and $U_{e3} = 0$,
while it naturally accomodates maximal amplitude $\nu_e$ oscillations
and a tiny $\Delta m^2_\odot$.
Maximal $\nu_\mu$--$\nu_\tau$ mixing follows from the reasonable assumption
that two Yukawa couplings are almost equal.
Neutrino masses are small because
they are radiatively generated at two-loops level.
Indeed,
the fact that two neutrino masses are as {\em large} as $0.06\, {\rm eV}$
practically forces the new mass scale,
at which the extra scalars lie,
to be in the $10\, {\rm TeV}$ range;
while deviations from $e$--$\mu$--$\tau$ universality
in $\mu$ decay and in $\tau$ decay
should be close at hand.
The model requires some physical mechanism
for generating a tiny soft breaking of $\bar L$.

\vspace*{5mm}

It is a pleasure to thank Walter Grimus for his generous help
and valuable advice,
and for reading and criticizing two drafts of the manuscript.
I also thank Evgeny Akhmedov for calling my attention
to a potential problem with the model,
and for reading the final version.

\section*{Figure captions}

\noindent Figure 1: Two-loops Feynman diagram which generates
${\cal M}_{e \mu}$.

\vspace*{1mm}

\noindent Figure 2: Two-loops Feynman diagram which generates
${\cal M}_{\mu \tau}$.

\vspace*{1mm}

\noindent Figure 3: One of the three-loops Feynman diagrams which generate
${\cal M}_{e e}$.

\newpage

\begin{figure}
\begin{center}
\setlength{\unitlength}{1.0bp}
\begin{picture}(340,527)
\put(122,488){$f^-$}
\put(212,488){$f^-$}
\put(175,468){$g^{2-}$}
\put(91,433){$\nu_{e L}$}
\put(124,433){$\mu_L$}
\put(124,423){$\tau_L$}
\put(151,433){$\mu_R$}
\put(151,423){$\tau_R$}
\put(178,433){$e_R$}
\put(207,433){$e_L$}
\put(236,433){$\nu_{\mu L}$}
\put(152,405){Figure 1}
\put(122,317){$f^-$}
\put(212,317){$f^-$}
\put(175,314){$g^{2-}$}
\put(175,285){$h^{2-}$}
\put(92,264){$\nu_{\mu L}$}
\put(124,264){$e_L$}
\put(151,264){$e_R$}
\put(178,264){$e_R$}
\put(206,264){$e_L$}
\put(235,264){$\nu_{\tau L}$}
\put(152,244){Figure 2}
\put(109,152){$f^-$}
\put(227,152){$f^-$}
\put(152,127){$g^{2-}$}
\put(58,93){$\nu_{e L}$}
\put(86,93){$\mu_L$}
\put(86,83){$\tau_L$}
\put(108,93){$\mu_R$}
\put(108,83){$\tau_R$}
\put(140,93){$e_R$}
\put(189,93){$e_R$}
\put(220,93){$\mu_R$}
\put(220,83){$\tau_R$}
\put(245,93){$\mu_L$}
\put(245,83){$\tau_L$}
\put(270,93){$\nu_{e L}$}
\put(166,75){$h^{2-}$}
\put(203,73){$g^{2-}$}
\put(152,53){Figure 3}
\put(0,0){\includegraphics{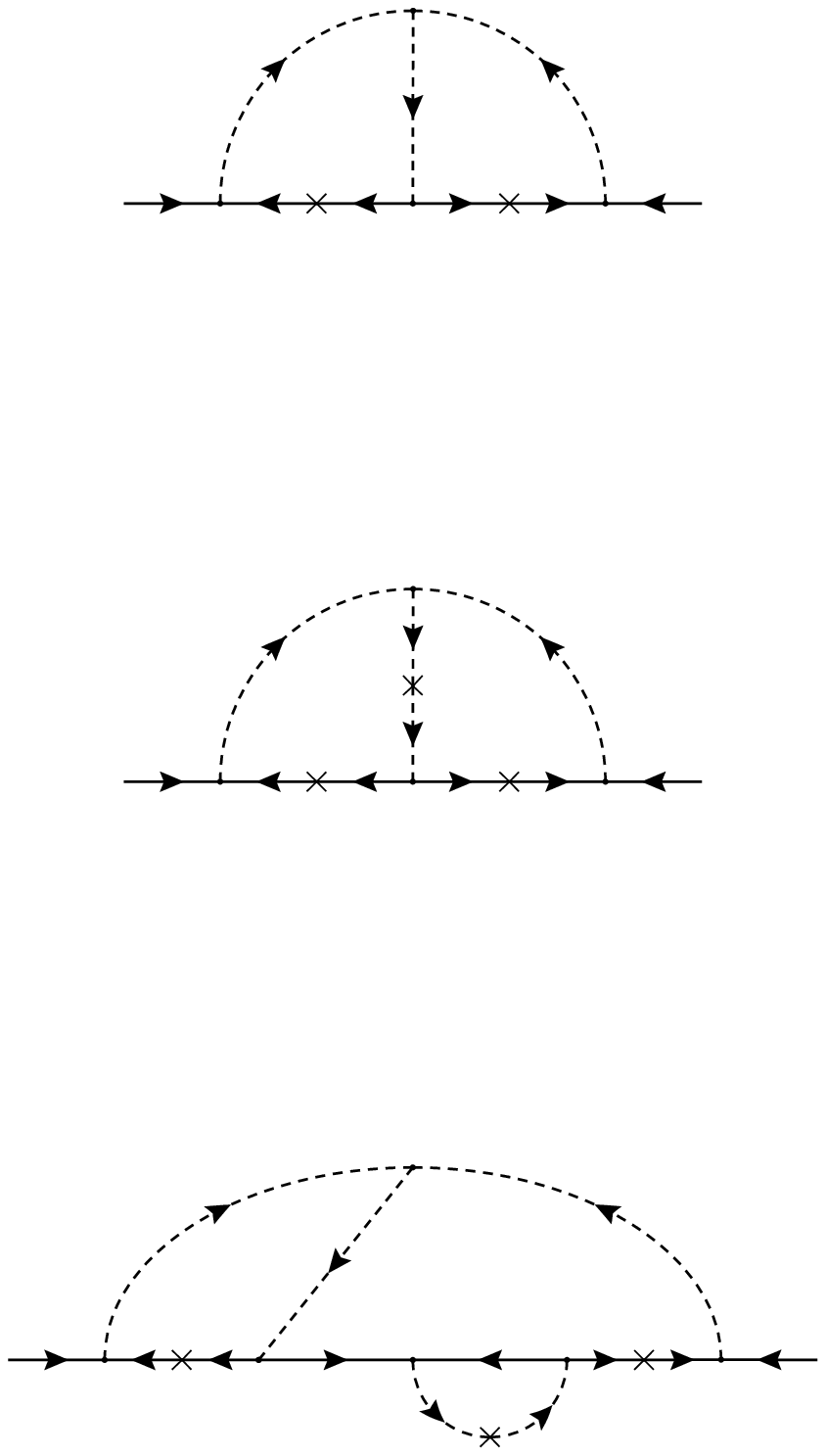}}
\end{picture}
\end{center}
\end{figure}

\end{document}